# Designing forecasting software for forecast users: Empowering non-experts to create and understand their own forecasts

*Completed Research Full Paper*

**Richard Stromer**
Stanford University
stromer@stanford.edu

**Oskar Triebe**
Stanford University
triebe@stanford.edu

**Chad Zanocco**
Stanford University
czanocco@stanford.edu

**Ram Rajagopal**
Stanford University
ram.rajagopal@stanford.edu

## Abstract

Forecasts inform decision-making in nearly every domain. Forecasts are often produced by experts with rare or hard to acquire skills. In practice, forecasts are often used by domain experts and managers with little forecasting expertise. Our study focuses on how to design forecasting software that empowers non-expert users. We study how users can make use of state-of-the-art forecasting methods, embed their domain knowledge, and how they build understanding and trust towards generated forecasts. To do so, we co-designed a forecasting software prototype using feedback from users and then analyzed their interactions with our prototype. Our results identified three main considerations for non-expert users: (1) a safe stepwise approach facilitating causal understanding and trust; (2) a white box model supporting human-reasoning-friendly components; (3) the inclusion of domain knowledge. This paper contributes insights into how non-expert users interact with forecasting software and by recommending ways to design more accessible forecasting software.

**Keywords**

Human-computer interaction, forecasting, design guidelines, software, accessibility, explainable AI.

## Introduction

To forecast is to predict the future, given what is known at present (Hyndman and Athanasopoulos 2018). Given the importance of this task for the functioning of society, forecasting is now a widely used tool for optimization and smarter decision-making. There are various application domains, such as sales, finance, transportation, energy, inventory, and logistics. The shift towards data-driven decision-making increases also the demand for better forecasting (Hosseini et al. 2022). The nature of forecasting is rapidly changing as datasets grow from dozens to millions of samples (Salinas et al. 2019).

Recent years have seen rapid innovation in forecasting methods with Deep Learning leveraging more data to achieve higher accuracy (Oreshkin et al. 2020). Their adoption accelerated with the proliferation of more affordable and accessible computing resources, leading to fully automated solutions (He et al. 2021). Nonetheless, there is an adoption gap from research to practice (Asimakopoulos and Dix 2013; Tashman and Hoover 2001a).

Modern forecasting generally does not include relevant human judgment nor domain knowledge, instead purely relying on data. Human judgment tends to be biased due to its subjective nature (Hyndman and Athanasopoulos 2018) and less accurate than statistical approaches (Carbone et al. 1983). However, better results are achievable by involving humans in data-driven forecasting (Arvan et al. 2019; Petropoulos et al. 2018). The inclusion of human judgement into modern forecasting methods is challenging and remains an open research area.

A forecast is subject to human perception and acceptance, as its usefulness depends on human trust and understanding (Lakkaraju et al. 2016). A lack of transparency may lead to rejection and a return to human judgement. Forecasts are mostly produced by forecasting experts or special software. End users, usually technical domain experts or non-technical managers, often lack the expertise to evaluate and adjust forecasts produced by advanced methods. This reduces the likelihood of acceptance and leads to a return to judgmental forecasting. As a result, advanced forecasting tools and the forecasts they produce are often inaccessible to forecast end users.

In this paper, we investigate how to design accessible, transparent, and relatable forecasting software for end users. We focus on integrating state-of-the-art forecasting methods with design aspects to empower non-expert users with tools they can operate. To gather feedback of novices and non-experts on the human-computer interaction (HCI) with forecasts, we prototype an end-to-end forecasting software with a graphical user interface. We analyze how design choices affect users' forecasting outcomes, what their expectations and challenges are, and what makes forecasting software useful to them.

Our results show three main considerations for making forecasting more accessible to a broader audience: (1) a safe stepwise approach allowing forecasting users to learn about the model's reaction to their action (causality); (2) a white box model turning the forecasting complexity into human-reasoning-friendly components; (3) the need for domain knowledge to include external factors in the model and evaluate and interpret forecasts for downstream tasks.

*Contribution.* Our prototype and user feedback provide insights into forecasting users' challenges and expectations. We emphasize the importance of interpretable models and domain knowledge in creating and accepting forecasts by non-experts. Additionally, we offer design recommendations for accessible forecasting software. The next section provides an overview of human-centered forecasting and related work, followed by the presentation of our prototype and user study design, discussion of relevant findings, and conclusion with implications and limitations.

## Human-centered forecasting

*Forecasting.* The goal of forecasting is to predict a variable's future value most accurately, given all the information available (Hyndman and Athanasopoulos 2018). The art of forecasting has been practiced by humans for thousands of years. Observations of natural phenomena were transferred into concepts following human reasoning. Over time these concepts were formalized and standardized to circumvent human limitations and to prevent human bias.

*Big Data and Deep Learning.* The steady growth in data availability brings a shift from single to collections of millions of time series (Salinas et al. 2019). Modern forecasting algorithms provide the technology to use increasing amounts of data and achieve higher accuracy (e.g., Deep Learning). However, these approaches often do not support the inclusion of real world and domain knowledge or are regarded as black-box models. Overcoming these constraints, which can often render resulting forecasts unusable for practitioners, is an open research area (Hosseini et al. 2022).

*Human Judgement and Domain Knowledge.* On the other extreme, judgmental forecasts are forecasts made by human experts considering specific domain knowledge and their subjective judgement of the situation. While human bias can lead to poorer performance compared to statistical forecasts (Goodwin et al. 2011), human judgement is still needed for manual adjustments and unprogrammable aspects (Fildes and Goodwin 2013). In recent years, literature has adopted a more positive perspective regarding judgmental forecasting, and human forecasters have been noted for improving model accuracy compared to pure statistical models (Goodwin et al. 2011; Petropoulos et al. 2018). Overall, literature recommends "blending judgement with statistical methods" (Lawrence et al. 2006) and argues the topic has been long overlooked by researchers and practitioners (Arvan et al. 2019).

*Forecasting Support Systems.* Forecasting Support Systems (FSS) are a special type of Decision Support Systems where forecasts are created based on the interaction between a forecaster and the information system (computer) (Fildes and Goodwin 2013). FFS tend to specialize in their application domain and are typically limited to import tasks, evaluation, presentation, and variable selection (Tashman and Leach 1991). How to embed modern forecasting methods and best practices remains in FSS an open research area.

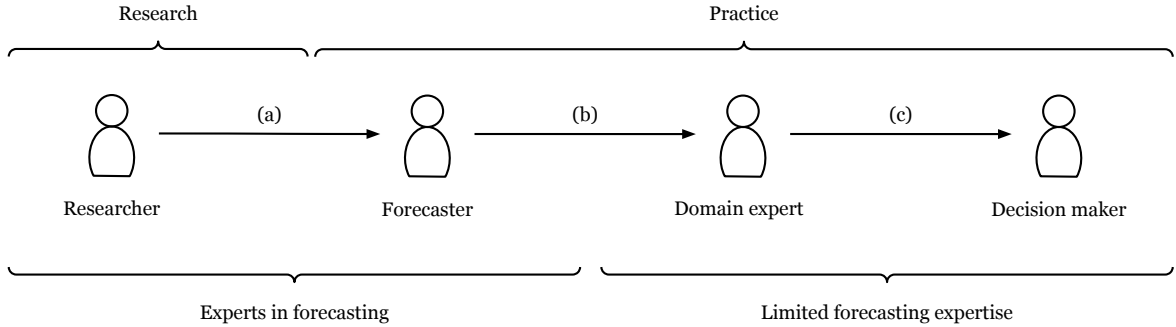

Figure 1: Roles in forecasting

*Who creates and who uses forecasts?* We can organize the people in forecasting in groups by their primary responsibility: Researchers, forecasters (expert practitioner), domain experts, and decision makers (see Figure 1). The interactions between the different roles can be summarized as: (a) a researcher is tasked with advancing the state-of-the-art (SOTA) of forecasting. Researchers generally communicate their innovation unidirectionally (Arvan et al. 2019). (b) Forecasting experts skillfully create and explain forecasts in practice, often communicating with a domain expert to embed know-how. (c) A domain expert seeks to understand and possibly improve the forecast with their domain knowledge, while a decision maker is responsible for the consequences of a particular decision informed by the forecast. A decision maker consults with the domain expert and forecaster, seeking to understand the context and trustworthiness of the forecast.

*Expertise gap.* Domain experts and decision makers are often not forecasting experts nor computer scientists. These expert skills are usually hard to acquire and a difficult barrier to overcome. We define this barrier as expertise gap. Without a forecaster's expertise the tools available and accessible to them might be limited to human judgement, experience-based heuristics, or linear models in spreadsheet applications. Ideally, forecasters and domain experts work closely together to create forecasts. Both are highly trained roles and rare, especially with the growing demand for forecasts. As there are not enough forecasting experts to overcome the expertise gap, we strive to make forecasting more accessible to a broader audience. We believe that interactive forecasting systems can help to democratize access to modern forecasting methods for non-expert forecasters like domain experts, decision makers and forecasting novices.

*Forecast acceptance.* We refer to forecast acceptance issues when a given forecast is rejected due to human trust and understanding issues. When a forecast was generated by a forecaster (expert) or a forecasting tool (software), the decision maker and domain expert have no direct ownership of the forecast. A widespread problem in this situation is that the decision maker and domain expert do not sufficiently understand or trust the forecast (Lakkaraju et al. 2016). In practice this results in the forecast being discarded (Asimakopoulos and Dix 2013). Despite its significance, the acceptability of forecasting software is an understudied research area (Arvan et al. 2019).

*Accessibility.* For forecasting tools to be accessible to non-expert forecasters, both the expertise gap and forecast acceptance need to be addressed.

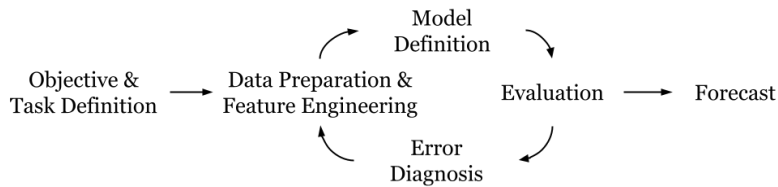

Figure 2: Forecasting workflow overview

*Human Computer Interaction.* Humans interact with forecasts often through computers, in two ways:

(1) Human input: Humans define the objective and context for the forecasting model by infusing their domain knowledge, judgement, assumptions, and modeling decisions (Önkal et al. 2009).
(2) Human perception: Humans receive information from forecasts through interpretation, evaluation, understanding, and trust (Hosseini et al. 2022).

This two-fold interaction is always subject to human perception and contextual factors, which can cause interruptions and misconceptions (Lakkaraju et al. 2016). Historically, there has been a lack of attention on the usability and human interaction with forecasting software (Tashman and Hoover 2001b).

### *Related work*

*Human Computer Interaction.* A variety of HCI studies explore forecasting related topics, including specific forecasting workflow steps, such as feature selection (Krause et al. 2014) and model selection (Sun et al. 2020); trust and explainability in specific application domains (Hohman et al. 2019; Ma et al. 2022; Saluja et al. 2021). Few studies cover the full forecasting workflow, but focus on experts only (Hohman et al. 2019), have no user interface (Lakkaraju et al. 2016), or have yielded no conclusive results (Ehsan et al. 2021). The "Human in the loop" concept was re-popularized by (Taylor and Letham 2017) as an iterative forecasting process. An overview of HCI studies regarding Explainable AI is provided in (Rong et al. 2022).

*Explainable Artificial Intelligence (XAI).* Explainability in time series is still relatively unexplored (Saluja et al. 2021). Explaining the inner workings of forecasting models is significantly more challenging when utilizing Deep Learning methods. LIME (Ribeiro et al. 2016) and SHAP (Lundberg and Lee 2017) are two popular model-agnostic methods for explaining the importance of variables. Other efforts focus on learning explanatory auxiliary models (Zhang et al. 2019). A drawback of such post-hoc methods is that they add an additional layer of complexity. Inherently interpretable forecasting models, such as hybrid forecasting methods, aim to bridge the gap between black-box deep learning methods and simple traditional forecasting methods (Hosseini et al. 2022; Oreshkin et al. 2020; Triebe et al. 2021). However, they still require a certain level of expertise to operate well.

*Automated forecasting systems.* Using computers to assist humans in forecasting through automation of calculations and statistics is no new practice (Fildes and Goodwin 2013; Tashman and Leach 1991). A large part of forecast automation literature only covers a specific subset of the steps in a forecasting pipeline and notes that future research on holistic automation of the entire workflow is necessary. (Meisenbacher et al. 2022) provides an overview of the forecasting workflow and automation methods.

Efforts are underway to better understand the human aspects of forecasting as the development of modern methods continues. Yet more research is needed to understand the full forecasting workflow through a human-computer interaction (HCI) perspective to improve accessibility, trust and understanding.

## Prototype

Intending to take a practical step toward making forecasting tools more accessible, we designed a prototype accessible to users without forecasting or programming skills. The prototype (see Figure 3) is designed so domain experts and decision-makers can interact with a forecasting model through a graphical user interface (GUI). Our design aims to balance automation, transparency, and user control.

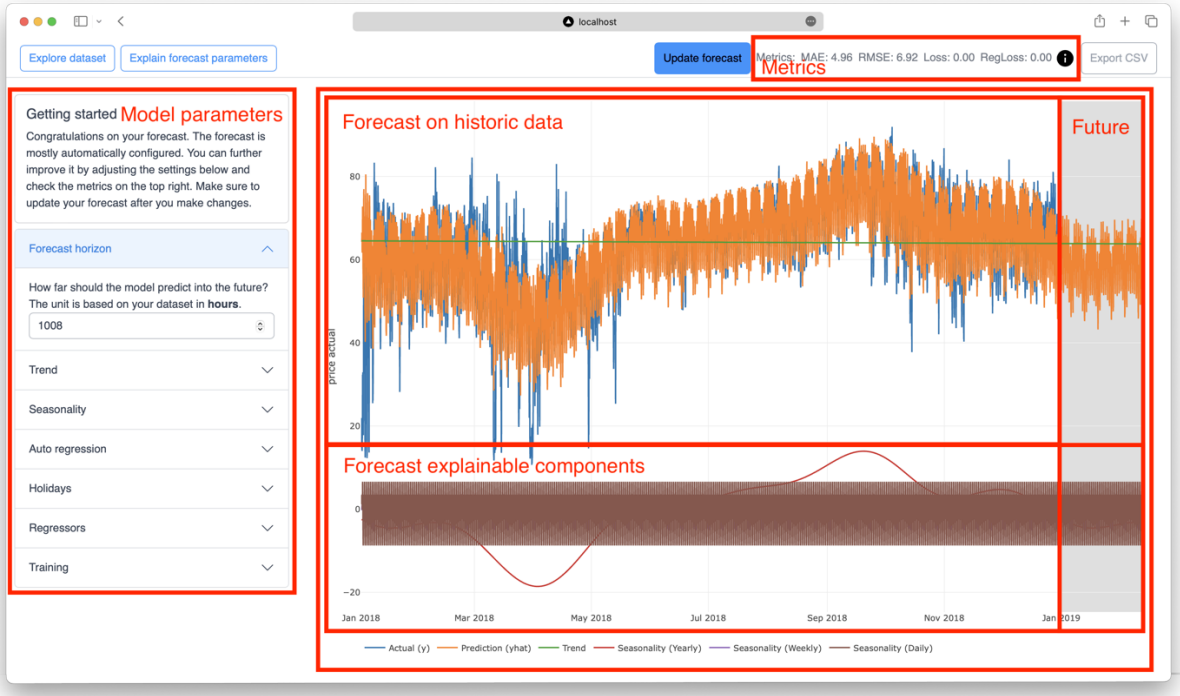

Figure 3: Our prototype with the key functionalities highlighted.

*User journey.* The user journey of our software prototype (Figure 4) follows the forecasting workflow (see Figure 2): Users begin this journey by importing their dataset from a source. On the next screen, they then define the time and forecast target fields for their imported dataset. Optionally, they can select a screen that allows them to explore, validate and visualize the dataset in depth. Moving on to the main screen, users receive an initial forecast based on a baseline model definition. Within this same screen, users can adjust the model configuration while visually inspecting the forecast and tracking metrics. The main screen workflow allows for iterative improvements. In addition, users can enter a screen that explains the model more transparently and provides a decomposition of the forecast. At this point in the workflow, users can choose to return to the dataset explorer to make any additional refinements. Finally, users can export their forecasting results as a spreadsheet or figure for later applications.

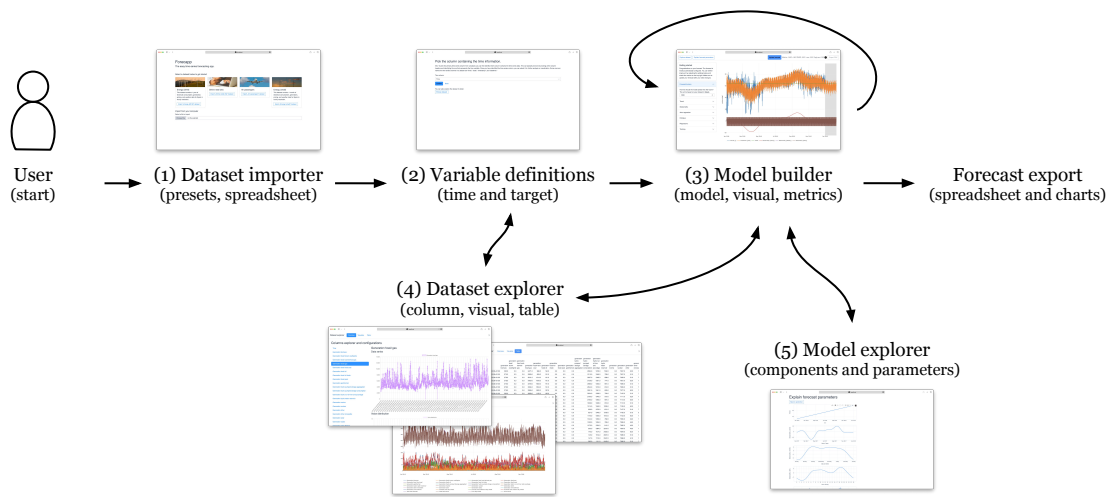

Figure 4: The user journey covering data import, definitions and model building.

*System architecture and technology.* The prototype is a web application with a graphical user interface. The user configures the model by embedding their domain knowledge and receives a forecast with interpretable components and metrics. At the core of our prototype is the modern hybrid forecasting model NeuralProphet (Triebe et al. 2021) which enables a stepwise human-in-the-loop workflow. The front-end is built in TypeScript using React, Redux, and NextJS. The back-end is written in Python with FastAPI and NeuralProphet.

## Method

To gather feedback about our forecasting software prototype, we conducted a user study (n=19). Participants engaged in a think-aloud study and a semi-structured interview. Earlier versions of the design process included a click dummy to validate the user journey and a technical prototype. This way, we collected qualitative data about user interaction with the prototype while providing context on their forecasting needs, use cases, domain expertise, and trust. In addition, we measured user perception in quantitative terms with the System Usability Scale (SUS) (Brooke 1996).

*Participants.* The 19 participants were recruited using professional and academic contacts. The median forecasting experience across all participants was one year, with 6 having no prior forecasting experience, and another 4 having less than a year of experience. The maximum experience was three years among those creating forecasts as part of their research or work duties. Participants with more advanced coding skills were also more knowledgeable about machine learning.

*Study design.* For our user study, participants were given an imaginary use case of predicting the energy price for the next month. They were provided with a dataset for Spain containing hourly data for electricity demand, generation, price and related weather data from 2015 to 2018 (4 years). We combined qualitative methods of observing the user interactions in think-aloud studies and with quantitative methods, namely the participant's best model score, a questionnaire for assessing forecasting expertise level, and the System Usability Scale (Brooke 1996). The goal was to identify usability issues, missing functionalities, and validate the workflow. The semi-structured interviews were designed to provide context and insights into (1) needs and use cases, (2) readiness and completeness of the tool for regular use, (3) general feedback on the current prototype, (4) knowledge, skills and learnings, and (5) confidence and trust in the model and forecasting results.

*Data Analysis.* We recorded and transcribed all interviews, each lasting 27 minutes on average (range 15 - 50 minutes). All personal information was removed, and we processed only anonymized data. In total, we generated 440 observations from these transcripts, including statements and user quotes, that we then iteratively labeled by emergent theme. The labeled observations were transferred to an affinity diagram to further identify common patterns (Lucero 2015). Similar and repetitive observations were flagged to indicate higher relevance. The context of all observations was preserved with anonymized participant numbers, the transcript timestamp, and the exact version of the prototype iteration.

## Results

We identified the following results: (1) A step-by-step forecasting approach helps domain experts generate their own forecasts and understand cause and effect relationships; (2) a white-box model consisting of explainable components is important both for the effective embedding of domain knowledge and for building understanding and trust into the produced forecasts; (3) domain expertise empowers users to operate forecasting software and has the potential to improve decision-making. We used the SUS as a proxy to further contextualize the participants' statements during the user study. The SUS scores range between 12 and 62, out of a potential maximum of 100. The Figma iterations yielded overall higher SUS scores.

*Stepwise approach.* Users adjusting one parameter at a time managed complexity better and developed a deeper understanding of the system dynamics. They understood causality by observing the model's reaction to their actions: "I think the step-by-step process is good." [P5] This also is reflected by P7 achieving the best RMSE of 6.81 among our participants with their stepwise approach. The associated sense of personal contribution builds trust: "I feel like the process gave me confidence when I saw error values decreasing" [P7]. We also found metrics improved user motivation by providing them with a measure for making progress. Multiple users asked for a version history to keep track of changes (P11, P12, P13, P14, P17, P19). In this respect, users may be more willing to try new approaches with the knowledge that any changes can be easily reverted, as in the words of P19: "I didn't remember my error and my deviation [metrics] from time to time. So, it would be quite nice if I had [...] something like a history: What did I change? What was the outcome?".

*Explainable model composition.* We found using a white box model composed of human-reasonable components helps build understanding of forecast dynamics and confidence in outcomes. Time-series data often has inherent complexities. While it is challenging to integrate these complexities in a model, it is even harder to attribute them correctly using human understandable concepts. We found most users quickly understand widely used terminologies such as trend, seasonality, and holidays in the context of forecasting. These analogies helped them to draw specific insights and reason with the results by validating components independently. For example, P9 could quickly follow and interpret the patterns of daily seasonal differences in energy pricing with their domain knowledge: "[Energy price] goes up at the start of the week and then kind of down. That pattern is pretty consistent. Daily seasonality has like a peak before noon and one peak in the afternoon. That makes sense, I suppose, when people are working and then using a lot later." P15 further interpreted and compared the seasonality with historic data: "The weekly [seasonality] is obvious. OK, that is very good. That is very well detected".

*Domain knowledge.* Incorporating domain knowledge improved forecasting outcomes, yet in practice it was challenging to achieve due to differences in terminology used in forecasting and application domains. To add external factors to the model, users needed to have knowledge about domain specific causal relationships (e.g., weather with sun hours influences solar energy production). Domain experts had their own relevant factors for the model and were actively searching for integration options. In contrast, users without domain knowledge struggled to complete the tasks, felt more uncertain and could not elaborate on what insights they might possibly generate. For example, P18 expressed they are "not feeling familiar in this field, not sure if I would know how to make it learn". This was also shown for users who stopped interacting with the software after an initial exploration, as soon as they had no additional domain knowledge to integrate. Validating the results and checking their plausibility also requires domain knowledge, as users were otherwise limited in their evaluation to comparing the visual fit and the metrics. P8 stated, "It means the prices increase, and then they decrease again. And I don't know if that makes sense with energy prices". However, we found that providing contextual explanations and guides helped mitigate the users' difficulties in understanding forecasting specific language and abbreviations. Many users were happy to learn while using the application, and having resources in immediate proximity for when users encountered unfamiliar terminology was important: "It all made sense to me with the kind of hints there" [P17]. Moreover, it also became evident that forecasting developers need to balance the abstraction of terminologies. Some users were overwhelmed with technical functionalities with P15 mentioning, "I see advanced options, OK, yeah, I'm not going to touch that one".

*Feature-related results.* Users expressed the need for further improvements around four areas:

(1) *Data preparation:* Users asked about the import format, meta information (descriptions, location, columns, units), validation (missing values, non-values, duplicates, resolutions), analytics (correlation charts, histogram binning), and transformations (aggregations, merging, mathematical operations).
(2) *Visualization:* Large differences in visualization expectations were influenced by users' background, e.g., TradingView (finance), Plotly (data science), or Financial Times (economists). We found that visualizing multi-step forecasts was more confusing than helpful for novice users. Flexible navigation, zooming and variable filtering in the visualization were considered by all users as key features.
(3) *Evaluation and metrics:* Initially, users did not understand the abbreviated metrics: "I'm not entirely certain what they mean." [P10] or "Yeah, it gives me a score here. It's not clear what it is." [P3]. With tooltip explanations this issue was resolved: "OK. That is nice to see the metrics up here" [P14]. The absolute metric values were critiqued for lacking a qualitative interpretation, like 'good' or 'bad'. Beyond judging the model quality by its fit to historic data, users were interested in its performance relative to other models.
(4) *Automation:* Users expressed the desire to automate the optimization of parameters. The automatic value selection for machine learning parameters (epochs, learning rate, batch size) received positive user feedback, and they asked to extend this feature, e.g., for autoregression. More knowledgeable users were more vocal about automatic optimizations to gain better results without manual intervention.

# Discussion

Our analysis of novice and non-expert forecasting user feedback yielded three key considerations to make forecasting more accessible for a broader audience: (1) provide users with a safe stepwise approach to learn how the model reacts to their actions (causality); (2) use a white box model breaking down complexity for human friendly reasoning; (3) the essential role of domain knowledge for both operating the model and interpreting its results for downstream tasks. We believe our contributions can help guide designers and developers in creating more accessible forecasting applications.

The results from our user study provide insights into the challenges and expectations of our non-expert user group. Our main results are in alignment with the findings of existing literature and provide more detailed and human-centered insights. The stepwise forecasting approach some of the non-expert users discovered matches the iterative nature of the forecasting workflow overall. The white box model decomposition with interpretable components might have its origins in separating trend and seasonalities to achieve stationary time series.

Based on our findings we propose these recommendations for designing accessible forecasting systems:

(1) *Forecasting playground:* Create a safe environment for beginners to experiment and learn. Make sure they can make 'quick wins' and achieve at least some of their objectives easily. Allow them to go into detail at a self-defined pace and ensure the system remains consistent and functional.
(2) *Educate and explain:* Teach beginners about forecasting on a voluntary basis. Provide them with recommendations on how to approach the forecasting workflow in general. Explaining unfamiliar words and concepts should be included in the context of the software application. Forecasting knowledge helps users with understanding and this way they can dive deeper later.
(3) *Forecasting best practices:* Help users incorporate best practices by providing recommendations and guidance, e.g., help them create a stationary time-series by removing trend and seasonality. Keeping the user in the loop allows them to remain in control and become more knowledgeable.
(4) *Domain specific questions:* Asking users relatable questions in their domain-specific language helps make the software more accessible, improving user understanding and enabling users to provide more specific answers, for example: "Which marketing campaigns did you run for your shop last year?"

Our findings suggest that forecasting software for professionals and experts might benefit from better accessibility and being more inclusive to a broader audience. In practice there are several other factors, which further limit the design choices of such forecasting software, e.g., economic considerations, technical constraints, and domain-specific conventions and requirements.

*Limitations.* Building generic forecasting software is challenging due to the high level of heterogeneity and complexity across different domains and datasets. While extending the prototype with domain specific forecasting templates and guidelines would have been an important addition, the needs and expectations varied too much across domains. This trade-off between domain specific fit and domain-agnostic generality will remain a delicate balance. Limitations in depth of the prototype include the absence of a model selection feature, making use of NeuralProphet's interpretable components. Model selection might allow for better performance and is one example of the further complex forecasting workflow, not being fully represented. Our prototype is intentionally limited for catering to novice and non-expert users.

*Future work.* To extend our work, another iteration of the forecasting software would help to address user feedback and add the most requested features. This includes more powerful earlier workflow steps like data selection, collection, validation and transformation for custom datasets, the integration of uncertainty visualization and probabilistic forecasting, and adding version control of the parameters, metrics and model history. Furthermore, the integration of feature decomposition, providing suitable model selection functionalities and advanced feature engineering would be relevant. To create better forecasting software, we need to gain a more holistic view of the industries and domains utilizing forecasts, in particular their application scenarios and use cases. This improved understanding of goals, tasks and existing solutions allows us to transfer forecasting advancements and accessibility faster across domains and industries.

*Outlook.* Forecasting impacts decision-making in every aspect of modern society, from daily weather updates to economic forecasts for companies, organizations, and governments. Human-in-the-loop forecasting is essential, as it integrates domain knowledge and judgement into the forecast objective, leading to better understanding and acceptance of results. Accessible technologies, like ChatGPT and Stable

Diffusion, have demonstrated the potential of harnessing complexity through easily accessible user interfaces. We believe that making forecasting more accessible can enable better decision-making, increasing resilience in a rapidly changing world.

# Conclusion

We developed a prototype of a forecasting tool for non-experts and collected feedback from 19 users to gain insights into their needs and expectations. Our findings suggest encouraging users to make iterative stepwise adjustments and white-box models with human-interpretable components improve accessibility. We encourage future research building on our results in human-centered forecasting especially with respect to accessibility and trust. Improved accessibility can better leverage domain expertise for more accurate forecasts, with potential wide-ranging benefits to various domains such as sales, climate, risk planning, production, electricity grids, and beyond.